# Detailed structural and topological analysis of SnBi$_2$Te$_4$ single crystal

Ankush Saxena[1,2], N. K. Karn[1,2], M. M. Sharma[1,2], and V.P.S. Awana[1,2,*]

[1]*Academy of Scientific & Innovative Research (AcSIR), Ghaziabad 201002, India*
[2]*CSIR National Physical Laboratory, New Delhi 110012, India*

**Abstract**

We report herein the successful synthesis of the topological material SnBi$_2$Te$_4$ in single-crystal form. Phase purity and unidirectional growth are evident from X-ray diffraction (XRD) patterns acquired from a powdered sample and a crystal flake. The crystalline morphology has also been visualized by acquiring a field-emission scanning electron microscope (FESEM) image. The crystal has been thoroughly characterized by means of Raman spectroscopy and X-ray photoelectron spectroscopy (XPS) measurements. The topological properties of SnBi$_2$Te$_4$ have been probed through magneto-transport measurements. SnBi$_2$Te$_4$ has been found to exhibit a small but non-saturating magneto-resistance (MR) up to ±12 T. The low-field magnetoconductivity (MC) of SnBi$_2$Te$_4$ at 2 K can be well explained through the Hikami–Larkin–Nagaoka (HLN) formalism, which confirms the presence of a weak anti-localization (WAL) effect in its crystal. Moreover, the non-trivial topological character has been evidenced through first-principles calculations using density functional theory (DFT), with and without spin-orbit coupling (SOC) protocols. A significant change in the bulk electronic band structure is observed upon the inclusion of SOC parameters, signifying topological properties of SnBi$_2$Te$_4$. Its topological non-trivial character has also been verified through calculation of Z2 invariants and the surface states spectrum in the (111) plane.

## Introduction

The quest for quantum materials with novel electronic states is a hot research topic in condensed matter physics. In this regard, topological insulators (TIs) are the most studied quantum materials, showing interesting electronic properties. An ideal TI shows conducting surface states accompanied by an insulating-type band gap in the bulk. The conducting surface states are topologically protected through time reversal symmetry (TRS) [1-3]. However, the most studied TIs, namely $Bi_2Se_3$, $Bi_2Te_3$, and $Sb_2Te_3$, do not show fully insulating bulk properties, and their electrical conductivity has some contribution from the bulk state. It is important to experimentally control the conductivity from surface and bulk states independently. In this regard, several methods have been developed, which include the synthesis of mixed topological insulators and pseudo-binary topological materials [4-9]. The topological pseudo-binary compounds have the general formula $A^{IV}B^{VI}$–$A^V_2B^{VI}_3$, where the superscripts denote the respective groups of the elements in the Periodic Table. Some of these compounds, namely $MnTe$-$Bi_2Te_3$ and $FeTe$-$Bi_2Te_3$, also show topological properties with intrinsic magnetic ordering [10-13]. These are considered as examples of magnetic topological insulators. Other materials in this family, such as $SnBi_2Te_4$ [14, 15], $SnSb_2Te_4$ [16], $GeBi_2Te_4$ [17], $GeSb_2Te_4$ [18], and $PbBi_2Te_4$ [7, 15], are non-magnetic and have recently been studied with regard to their non-trivial topological properties.

Angle-resolved photoelectron spectroscopy (ARPES), scanning tunnelling microscopy (STM), and transport measurements have been used to probe the surface states and insulating bulk properties of TIs. Recently, $SnBi_2Te_4$ has been found to show topological surface states in ARPES measurements and is considered to be a 3D-TI [19]. Topological surface states can also be probed through magneto-transport measurements. Generally, topological surface states are associated with a π Berry phase, which further leads to the observation of large anomalous Hall conductivity, quantum oscillations, and weak anti-localization (WAL) effects in magneto-transport measurements [20-22]. The WAL effect can be probed by analyzing the low-field magneto-conductivity (MC) of a TI. Materials of the general formula $A^{IV}B^{VI}$–$A^V_2B^{VI}_3$, namely $SnSb_2Te_4$, $PbBi_2Te_4$, and $SnBi_2Te_4$, have been found to show topological effects in magneto-transport measurements [14-16]. We are aware of two experimental reports in the literature, in which the surface states in $SnBi_2Te_4$ have been probed by magneto-transport measurements [14, 15]. In these reports, $SnBi_2Te_4$ was shown to have a π Berry phase, as manifested in a Shubnikov–de Haas (SdH) effect and a WAL effect [14, 15]. $SnBi_2Te_4$ shows a structural phase transition with the application of pressure, which is found to induce superconductivity therein [23].

In this work, single-crystal $SnBi_2Te_4$ has been grown by a simple solid-state reaction route. The phase purity of the synthesized $SnBi_2Te_4$ has been verified through XRD, FESEM, energy-dispersive X-ray spectroscopy (EDS), and Raman spectroscopy measurements. XPS measurements have also been performed to probe the chemical environments of the atoms in $SnBi_2Te_4$. Metallic behavior of the synthesized sample has been evidenced by ρ-T measurements. Electrons have proved to be the dominant carrier type in the synthesized $SnBi_2Te_4$ single crystal. Magneto-transport measurements show a V-type cusp at low magnetic field, characteristic of a WAL effect in topological insulators. This WAL effect has been confirmed by probing the low-field MC of $SnBi_2Te_4$ using the HLN formalism. This report, along with previous reports on the MC of $SnBi_2Te_4$, confirms the existence of topological surface states. These surface states are also predicted by DFT calculations, and show band inversion with inclusion of spin-orbit coupling (SOC) parameters. Z2 invariants have also been calculated, which show the presence of weak topology in the system, and this has been further verified by the surface-state spectrum of $SnBi_2Te_4$.

**Experimental**

High-quality (purity > 4N) Sn, Bi, and Te were combined in the requisite stoichiometric ratio. The mixture was ground under argon atmosphere in an M. Braun glovebox. It was then pelletized and encapsulated in a quartz ampoule at a pressure of $5\times10^{-5}$ mbar. A Proportional-Integral-Derivative (PID) controlled muffle furnace was used to heat the ampoule to 890 °C at a rate of 120 °C/h. The encapsulated sample was maintained at this temperature for 4 h to ensure homogeneity of the melt. It was then cooled to 500 °C at a rate of 1 °C/h, maintained at this temperature for 150 h, and then allowed to cool naturally to room temperature. A schematic of the synthesis process for obtaining single-crystal $SnBi_2Te_4$ is shown in Fig. 1. The synthesized single crystal, as shown in the inset in Fig. 1, was found to be easily cleavable along the growth axis. X-ray diffraction (XRD) patterns of a single-crystal flake and gently crushed $SnBi_2Te_4$ powder were recorded on a Rigaku Mini-Flex II table-top diffractometer employing Cu-$K_\alpha$ radiation of wavelength 1.5418 Å. The powder XRD (PXRD) pattern was analyzed with Full Proof software. VESTA software was used to construct the unit cell of the synthesized single crystal of $SnBi_2Te_4$. The surface morphology was inspected using a JEOL JSM 7200F FESEM equipped with an attachment for EDS. Raman spectra were recorded with a Renishaw inVia reflex Raman microscope with a laser wavelength of 514 nm. The sample was exposed to laser irradiation for 30 s, keeping the power below 5 mW. A Quantum Design physical property measurement system (QD-PPMS) was used to perform magneto-transport studies by the four-probe method. ρ-T measurements were performed in the temperature range from 250 K to 2 K and ρ-H measurements were performed under a magnetic field of ±12 T.

Transverse resistivity vs. magnetic field, i.e., $\rho_{xy}$ vs. H measurements, were performed in van der Paw geometry under magnetic fields up to 5 T.

To study and understand the topological properties of the synthesized $SnBi_2Te_4$ crystal, first-principles simulations were carried out, and Z2 invariants were also calculated to categorize the topology present in the system. For this, density functional theory (DFT)-based first-principles calculations were executed in Quantum Espresso [24, 25] to obtain the bulk electronic band structure and projected density of states (PDOS). For the calculations, atomic positions in the primitive unit cell of $SnBi_2Te_4$ were considered. The Perdew–Burke–Ernzerhof (PBE) generalized gradient approximation (GGA) was used to account for the electronic exchange and correlation. The wavefunctions were expanded in a plane wave with Gaussian smearing of width 0.01 on a Monkhorst–Pack k-grid of 9×9×9. The electronic bands were calculated with and without the inclusion of SOC. For the band structure calculated without SOC, we used the Standard Solid-State Pseudopotential (SSSP) library. For band calculations with SOC, the PSEUDODOJO pseudopotential library was used, which incorporates full relativistic approximations. For convergence of the self-consistent calculation, the cut-off was $1.2 \times 10^{-9}$ Ry, and a charge cut-off of 320 Ry and a wavefunction cut-off 45 Ry were used. Since $SnBi_2Te_4$ contains van der Waals gaps between its layers, we also performed calculations at a higher level of theory, namely dispersion-corrected DFT-D3 [26]. The DFT-generated Bloch wavefunction was wannierized using WANNIER90 software [27]. The electronic bands without SOC were reproduced in the range ±2 eV by disentanglement of 22 bands with a tolerance of $10^{-10}$ a.u., and the wannierization convergence tolerance was $7 \times 10^{-9}$ a.u. The states of Z2 invariants were determined by the evolution of Wannier charge centers (WCCs) in Brillouin zone planes, which were sampled on a much denser grid of $81 \times 81 \times 81$. The surface spectral function was calculated using the iterative surface Green's function method [28, 29] along the (111) plane.

**Results and Discussion**

Fig. 2(a) depicts the XRD pattern acquired from a mechanically carved crystal flake of a synthesized $SnBi_2Te_4$ single crystal. The Rietveld-refined PXRD pattern is shown in Fig. 2(b), which confirms that the synthesized sample crystallized in a rhombohedral structure with space group *R*-3*m*. All peaks in the PXRD pattern could be well fitted with the applied fitting parameters, showing the absence of any impurity phase that may have arisen through the insertion of extra atomic layers. The quality parameter of fitting, $\chi^2$ (goodness-of-fit), was found to be 2.78, which lies in the acceptable range. The lattice parameters and atomic positions obtained from Rietveld refinement are given in Table 1 and Table 2, respectively. $SnBi_2Te_4$ crystallizes in a similar crystal structure with the same space group symmetry as its parent compound $Bi_2Te_3$. The only change that occurs due to insertion of the SnTe layer is an enhancement of the lattice parameter along the *c*-axis, as revealed through Rietveld-refined lattice parameters. This is also evident from the single-crystal XRD pattern featuring sharp high-intensity peaks along the (0 0 3n) direction, as shown in Fig. 2(a). This confirms that the sample had only grown along the *c*-axis. Interestingly, despite having the same space group, i.e. *R*-3*m*, the topological pseudo-binary compounds do not all share the same number of septuple layers in a single unit cell. MTIs such as $MnBi_2Te_4$ and $FeBi_2Te_4$ contain four blocks of septuple layers in their unit cells, which results in crystal growth along the (0 0 4n) plane [10-13]. In contrast, other pseudo-binary topological materials contain three blocks of septuple layers in their unit cells, which results in crystal growth along the (0 0 3n) plane. This was revealed through XRD patterns acquired from crystal flakes, which show high-intensity peaks along the growth axis i.e. (0 0 3n) for the synthesized $SnBi_2Te_4$ single crystal. The presence of three blocks of septuple layers in $SnBi_2Te_4$ is also evident from the unit cell structure constructed using VESTA software, as shown in the right-hand-side inset in Fig. 2. This shows that a layer of SnTe was successfully inserted into the unit cell of $Bi_2Te_3$.

The surface morphology of the synthesized $SnBi_2Te_4$ single crystal was visualized through acquiring an FESEM image, as shown in Fig. 3(a). Fig. 3(a) confirms that the synthesized sample had a typical terrace-type morphology, which is commonly shown by single-crystalline materials. The purity and homogeneity were further checked through EDS measurements, as shown in Fig. 3(b)–(e). Fig. 3(b) shows the EDS pattern of the synthesized $SnBi_2Te_4$ single crystal, in which all of the constituent elements are seen to be in the required stoichiometric ratio, signifying the purity of the sample. The homogeneity of the synthesized $SnBi_2Te_4$ single crystal was determined through respective EDS mappings of Bi, Te, and Sn, as shown in Fig. 3(c)–(e). All of the constituent elements were found to be evenly distributed, signifying the homogeneity of the sample.

The vibrational modes of the synthesized $SnBi_2Te_4$ single crystal were delineated by recording its Raman spectrum, as shown in Fig. 4. The Raman spectrum of $SnBi_2Te_4$ could be deconvoluted into four peaks using the Lorentz fitting formula. Raman modes are observed at $59\pm1$ cm$^{-1}$, $91\pm1$ cm$^{-1}$, $101\pm1$ cm$^{-1}$, and $131\pm1$ cm$^{-1}$, in accordance with previously reported values [30]. The vibrational modes of $SnBi_2Te_4$ differ from those of the parent compound $Bi_2Te_3$. $Bi_2Te_3$ generally shows three Raman modes above 50 cm$^{-1}$, corresponding to $A^1_{1g}$, $E^2_g$, and $A^2_{1g}$ [31]. Differences between the Raman spectra of $Bi_2Te_3$ and $SnBi_2Te_4$ arise due to the insertion of an extra SnTe layer into the $Bi_2Te_3$ unit cell, which results in the appearance of vibrational modes due to Sn-Te bonds. In $Bi_2Te_3$, the most stable middle atomic layer consists of Te(II) centers, whereas in $SnBi_2Te_4$ this middle atomic layer of Te(II) centers is replaced by a layer of Sn.

The abovementioned peaks correspond to three types of Raman modes, namely low-frequency, mid-frequency, and high-frequency modes. These modes are similar to those observed for $SnSb_2Te_4$, which were assigned as $A_g^1$, $A_g^2$, $E_g^2$, $E_g^3$, and $A_g^3$ [16]. The low-frequency modes ($E_g^1$, $A_g^1$) and high-frequency modes ($E_g^3$, $A_g^3$) are similar to the $E_g^1$, $A^1_{1g}$, $E_g^2$, and $A^2_{1g}$ Raman modes observed for $Bi_2Te_3$ [31]. Schematic illustrations of the origin of these vibrations are shown as an inset in Fig. 4. The mid-frequency modes, $E_g^2$ and $A_g^2$, arise due to insertion of the SnTe layer. In these modes, only the Te atoms vibrate, as shown in the inset in Fig. 4. The Raman peaks at $59\pm1$ cm$^{-1}$, $101\pm1$ cm$^{-1}$, and $131\pm1$ cm$^{-1}$ are thus assigned to the $A_g^1$, $E_g^3$, and $A_g^3$ modes of $SnBi_2Te_4$. The presence of a mid-frequency mode, that is, the $E_g^2$ mode at $91\pm1$ cm$^{-1}$, signifies the successful insertion of an SnTe layer into the $Bi_2Te_3$ unit cell.

Chemical analysis of the synthesized $SnBi_2Te_4$ single crystal was performed by XPS, taking the C 1s peak as a reference. XPS patterns were acquired in the regions corresponding to Te 3d, Sn 3d, and Bi 4f, as shown in Fig. 5(a), (b), and (c), respectively. The Lorentz fitting formula was used to fit these XPS peaks. Fig. 5(a) shows the fitted XPS peaks in the Te 3d region, which correspond to spin-orbit doublets of Te, that is, $3d_{5/2}$ and $3d_{3/2}$. The XPS peaks for Te $3d_{5/2}$ and Te $3d_{3/2}$ are seen at $572.35\pm0.01$eV and $582.74\pm0.01$ eV, respectively, slightly displaced from their respective standard values [32] due to the bonding of Te atoms with Sn and Bi atoms. XPS peak positions depend on the electronegativities of elements. In $SnBi_2Te_4$, the Te atoms are bonded to both Bi and Sn, but the electronegativities of Bi and Sn are almost the same, which results in the appearance of Te peaks in identical positions for both bonds. The separation between these two peaks is 10.39 eV, matching the standard value of 10.39 eV quoted in ref. [32]. Fig. 5(b) shows the XPS peaks of the spin-orbit doublet of Sn, that is, Sn $3d_{5/2}$ and Sn $3d_{3/2}$, at $485.37\pm0.02$ eV and $493.76\pm0.01$ eV, respectively. The separation between these peaks is 8.39 eV, comparable to the standard value of 8.41 eV quoted in ref. [32]. Two further peaks are observed at $486.75\pm0.01$ eV and $495.18\pm0.02$ eV, attributable to $SnO_2$, which was formed by surface oxidation of the sample. Fig. 5(c) shows the XPS peaks of the spin-orbit doublet of Bi, that is, Bi $4f_{7/2}$ and Bi $4f_{5/2}$, at $157.64\pm0.02$ eV and $162.94\pm0.01$ eV. The values are slightly different from the standard values of XPS peaks of elemental Bi due to the bonding to Te atoms. The separation between these XPS peaks is 5.30 eV, comparable to the standard value of 5.31 eV quoted in ref. [32]. The XPS peak positions for Te, Sn, and Bi are listed in Table 3.

The resistivity (ρ) versus temperature (T) plot of the synthesized $SnBi_2Te_4$ crystal is shown in Fig. 6. The resistivity is seen to decrease with decreasing temperature, showing the metallic nature of $SnBi_2Te_4$. The residual resistivity ratio (RRR) was evaluated as 1.74, which is low, but comparable to previously reported values for $SnBi_2Te_4$ [15]. The resistivity was fitted over the entire studied temperature range (250 K to 2 K) according to the following formula [15]:

$$\rho(T) = \rho_0 + A * exp\left[\frac{-\theta}{T}\right] + B * T^2 \qquad (1)$$

Here, $\rho_0$ represents the residual resistivity, which arises due to impurity scattering, while the exponential and the quadratic term correspond to electron-phonon and electron-electron scattering, respectively. The obtained values of the fitting parameters were $\rho_0$=0.53 mΩ·cm, A=0.176 mΩ·cm, θ=72.5 K, and B=1.02×10$^{-8}$ Ω·cm·K$^{-2}$. Hall measurements were carried out in van der Paw geometry to determine the type of charge carriers and their mobility in the synthesized $SnBi_2Te_4$ single crystal. Fig. 6(b) shows transverse resistivity ($\rho_{xy}$) vs. H plots at different temperatures, namely 250 K, 200 K, 100 K, and 50 K. The Hall coefficient ($R_H$) was determined according to the single-band model, whereby the slope of the linearly fitted $\rho_{xy}$ vs. H plot gives the value of $R_H$. The values of $R_H$ were found to be negative at all of the measurement temperatures, implicating electrons as the major charge carriers in $SnBi_2Te_4$, consistent with a previous report [15]. Carrier concentration and mobility were determined according to the relationships $R_H = \frac{-1}{n_e e}$ and $\mu_e = \sigma R_H$ respectively. Here, $n_e$ is the carrier density, $e$ is the electronic charge, and σ is the conductivity of the synthesized $SnBi_2Te_4$ single crystal. The obtained values of carrier density and mobility are given in Table 4. Carrier density and mobility were plotted against temperature, as shown in Fig. 6(c). The carrier density was found to increase with increasing temperature, as shown by the black line in Fig. 6(c). The variation in mobility with temperature is shown by the red line in Fig. 6(c), which shows a monotonic decrease as the temperature is increased.

Magneto-transport measurements were carried out in a field range of ±12 T at temperatures of 2 K, 5 K, 10 K, 50 K, 100 K, and 200 K. MR% was calculated according to the following formula:

$$MR\% = [\rho(H)–\rho(0)]/\rho(0) \times 100\% \qquad (2)$$

Here, ρ(H) denotes the resistivity in the applied field and ρ(0) denotes the resistivity in zero field. The calculated MR% in the field range ±12 T at 2 K is plotted in Fig. 7(a). A clear V-type cusp is observed in the low-field (up to ±2 T) MR% data, which signifies the possible presence of WAL in the synthesized SnBi$_2$Te$_4$ single crystal [33]. To determine whether the observed MR originated from topological surface states or bulk states, the high-field MR% data at 2 K were fitted with a power law, namely MR%=A*H$^\gamma$. The value of the exponent γ determines whether the electrical transport is governed by topological surface states or by bulk states. A linear dependence of MR% on applied field, i.e. γ=1, would indicate that transport is dominated by topological surface states, whereas a quadratic dependence of MR% on applied field, i.e. γ=2, would indicate a dominance of bulk states. An intermediate value of γ, i.e., $1 < \gamma < 2$, would indicate that the transport properties have contributions from both topological surface states and bulk states [34]. Here, the exponent γ was evaluated as 1.16, implying that the transport is governed by both bulk and surface conducting channels. A combined plot of MR% vs. H at temperatures of 2 K, 5 K, 10 K, 50 K, 100 K, and 200 K is shown in Fig. 7(b). MR% remains almost constant up to 10 K, but significantly decreases at higher temperatures. More interestingly, the V-type cusp observed at low temperatures also broadens at higher temperatures. This implies that the bulk states start to contribute more to the conduction mechanism at higher temperatures. To gain more insight into the conduction mechanism of the synthesized SnBi$_2$Te$_4$ single crystal, the low-field MC at 2 K was fitted with the HLN equation [35], which is expressed as follows:

$$\Delta\sigma(H) = \frac{-\alpha e^2}{\pi h}\left[ln\left(\frac{B_\varphi}{H}\right) - \Psi\left(\frac{1}{2} + \frac{B_\varphi}{H}\right)\right] \quad (3)$$

Here, Δσ(H) is the difference in MC, that is, [σ(H)–σ(0)], $B_\phi$ is the characteristic field given by $B_\varphi = \frac{h}{8e\pi l_\varphi^2}$, where $l_\phi$ is phase coherence length, and Ψ is the digamma function. The value of the pre-factor α, which is included in the constant term, gives information about the kind of localization present in the system. The presence of weak localization (WL) results in a positive value of α, while a negative value of α shows that a WAL effect is operative in the system. In topological materials, a WAL effect arises due to the presence of a π Berry phase. The pre-factor α attains a value of either –0.5 or –1 for a material with a π Berry phase [33]. The value of α also signifies the number of conducting channels present in the system, with –0.5 per conducting channel. Thus, α = –1 would indicate that two distinct topological surface states (TSSs) are present in the system, one at the top and the other at the bottom [33, 36]. An intermediate value of α between –0.5 and –1 would imply that bulk states contribute to the conduction mechanism and that both of the TSSs are connected through bulk states [37], that is, the conduction has contributions from both bulk conducting channels and TSSs. Here, the HLN-fitted MC at 2 K is shown in Fig. 7(c). The low-field MC of SnBi$_2$Te$_4$ at 2 K is well fitted by the HLN equation, with α = –0.01087. This negative value of α signifies that a WAL effect is operative in the system, and the deviation from the standard value of –0.5 indicates that the electrical transport is not solely governed by the TSSs in SnBi$_2$Te$_4$. The observed value of α is much smaller than the standard value of –0.5, consistent with some earlier reports on materials with low MR% [16, 38, 39]. There are some materials with high SOC in which WAL arises due to bulk states, for which α is very large, of the order of $10^5$ [40, 41]. In the present case, the observed value of α is very low, which shows that electrical conduction in SnBi$_2$Te$_4$ has contributions from both the bulk states and TSSs.

To better understand the character of the synthesized SnBi$_2$Te$_4$ crystal, bulk electronic band structure and PDOS were calculated by a DFT-based first-principles method. The PDOS shows the orbital contributions to the band structure in a very straightforward manner [42]. As inputs for band-structure calculations, unit cell parameters were taken from the Rietveld-refined structure of the synthesized SnBi$_2$Te$_4$. Calculations were performed both with and without SOC parameters in Quantum Espresso with the PBE exchange-correlation functional [43]. The k-path followed for the calculations was determined with the aid of the SeeK-path k-pathfinder and visualizer program [44], which yielded H0 → L → Γ → S0 as the optimal path. Fig. 8(a) shows this particular k-path, as marked in the first Brillouin zone. It has been found that higher levels of theory, such as DFT-D3 [45-47], MP2 [48], and B3LYP [49], give more accurate results. Since the crystal under study has van der Waals gaps between its layers, we also included the van der Waals correction, for which DFT-D3 calculation was performed [26]. The dispersion correction energy was evaluated as –1.84 eV. Fig. 8(b) shows the DFT-D3-corrected band structure without including SOC parameters. The DFT-calculated bulk electronic band structures excluding DFT-D3 correction, without SOC and with SOC, are shown in Fig. 8(c) and (d), respectively, and the corresponding PDOS are shown in Fig. 8(e) and (f). The electronic band structure is seen to be essentially the same with or without DFT-D3 correction. The only difference is that the Fermi level is shifted, but it does not cross the bands in either case. The energy gap between the conduction and valence bands remains the same, but the Fermi energy changes. However, the Fermi energy still remains in the same gap. Thus, DFT-D3 correction did not affect the topology of the system. In the PDOS, the total DOS with SOC has a finite value at the Fermi level, indicating semi-metallic behavior. Flat-band analysis of wannierized bands indicates that the bands near the Fermi level have major contributions from the s orbitals of Sn and Te, and that the p orbitals of all three types of atoms contribute. These findings concerning the orbital contributions were corroborated by the PDOS calculations. No band crosses the Fermi level in the calculation without considering SOC bands; the valence and conduction bands remain well separated. The PDOS calculated without SOC and band structure also confirm an indirect 350 meV band gap. When SOC is included, the indirect band gap decreases markedly to 50 meV. Based on the band gap, the first impression is that $SnBi_2Te_4$ is a semiconductor, but further analysis indicates the existence of topological surface states.

Topological invariants characterize the topological properties inherent to a topological material. These topological invariants conform to the symmetry imposed by the material. The observed splitting of bands with inclusion of SOC parameters in the bulk electronic band structure, as well as the invariance of bands for time reversal momenta (±k), suggests that the system respects the TRS and these kinds of topological systems can be characterized in terms of Z2 invariants [50]. Furthermore, the space group of $SnBi_2Te_4$ is centrosymmetric *R-3m* (space group no. 166), and the symmetry-based indicator is Z2 [51-53]. Due to the symmetry present in the system, the Wannier charge evolution is shown only in the z-plane to compute the Z2 invariant. Here, we follow the Soulyanov–Vanderbilt [50] method of WCCs, calculated from maximally localized Wannier functions (MLWFs), to calculate Z2 invariants. The topological Z2 index is represented by four parameters ($\upsilon_0$; $\upsilon_1$ $\upsilon_2$ $\upsilon_3$), where $\upsilon_0$ represents the strong index and $\upsilon_1$, $\upsilon_2$, and $\upsilon_3$ represent the weak index. The weak index comprises the values of Z2 numbers for $k_i = 0.5$ planes (i = x, y, and z). Fig. 9 shows the evolution of WCC in the $k_z$ plane. Here, for the bands with SOC, we find that for all three planes, $\upsilon_0 = 0$, but $\upsilon_1 = \upsilon_2 = \upsilon_3 = 1$, indicating a topologically non-trivial state with weak topology in the system. Here, the weak index has no redundancy, and thus the Z2 index is (0; 1 1 1) for the studied $SnBi_2Te_4$ system.

The surface state spectrum was also computed using the iterative Green's function implemented in Wannier Tools. The surface card was set as (111), and the calculation was performed by taking 81 slices of one reciprocal vector. The k-path taken for the surface state spectrum calculation was L → Γ → L⁻, as shown in Fig. 10(a). Fig. 10(b) shows the plane in which the path lies. Fig. 10(c) shows the obtained spectrum, in which a possible Dirac cone is observable at the Γ point at energy around 0.09 eV. A similar Dirac cone of type II is observable very close to the Fermi level in the path L → Γ point, but not at the high-symmetry point. Both the surface state spectrum and the bulk electronic band structure indicate that the studied $SnBi_2Te_4$ crystal is a topological material.

**Conclusion**

In summary, an $SnBi_2Te_4$ single crystal has been grown by an optimized heat-treatment method. The crystallinity and purity of the synthesized crystal have been established through XRD, FESEM, Raman, and XPS measurements. Detailed Hall measurements have shown electrons to be the dominant charge carriers in $SnBi_2Te_4$ single crystal, showing it to be an n-type conductor. Magneto-transport measurements have indicated the presence of a WAL effect in $SnBi_2Te_4$, which has been verified by HLN modeling of low-field MC data at 2 K. The WAL in $SnBi_2Te_4$ has contributions from both bulk states and topological surface states. The presence of non-trivial band topology in $SnBi_2Te_4$ has been evidenced through bulk electronic band structure calculations, and further verified by topological non-trivial Z2 invariants. To the best of our knowledge, this is the first detailed investigation of Z2 invariants in $SnBi_2Te_4$, and has shown the presence of weak topology in the system. The presence of a Dirac cone in the surface space spectrum also confirms $SnBi_2Te_4$ to be a topological material.

**Acknowledgement**

The authors would like to thank the Director of the NPL for his keen interest and encouragement. Ankush Saxena would like to thank the DST for a research fellowship. M.M. Sharma and N.K. Karn would like to thank the CSIR for research fellowships. Ankush Saxena, M.M. Sharma, N.K. Karn are also thankful to the AcSIR for Ph.D. registration.

**Table 1**. Unit cell parameters obtained from Rietveld refinement of the PXRD pattern of the synthesized $SnBi_2Te_4$ single crystal.

| Cell Parameters | $SnBi_2Te_4$ |
|---|---|
| | |

| structure | rhombohedral |
|---|---|
| space group | *R*-3*m* |
| *a* | 4.395(4) Å |
| *b* | 4.395(4) Å |
| *c* | 41.606(2) Å |
| α | 90° |
| β | 90° |
| γ | 120° |

**Table 2**. Atomic positions of constituent elements of the synthesized SnBi$_2$Te$_4$ single crystal.

| Atoms | *x* | *y* | *z* |
|---|---|---|---|
| Sn | 0.0000 | 0.0000 | 0.428(1) |
| Bi2 | 0.0000 | 0.0000 | 0.428(1) |
| Bi1 | 0.0000 | 0.0000 | 0.500(0) |
| Te2 | 0.0000 | 0.0000 | 0.288(3) |
| Te1 | 0 | 0 | 0.135(9) |

**Table 3**. XPS peak positions and full-widths at half-maximum (FWHM) of constituent elements of the synthesized SnBi$_2$Te$_4$ single crystal.

| Element | Spin-orbit doublet | Binding Energy (eV) | FWHM (eV) |
|---|---|---|---|
| Sn | 3d$_{5/2}$ | 485.37±0.02 | 0.97±0.06 |
|  | 3d$_{3/2}$ | 493.76±0.01 | 0.62±0.07 |
| Bi | 4f$_{7/2}$ | 157.64±0.02 | 0.83±0.01 |
|  | 4f$_{5/2}$ | 162.94±0.01 | 0.77±0.02 |
| Te | 3d$_{5/2}$ | 572.35±0.01 | 1.10±0.01 |
|  | 3d$_{3/2}$ | 582.74±0.02 | 1.11±0.02 |

**Table 4**. Parameters obtained from Hall measurements.

| Temperature (K) | Hall Coefficient, R$_H$ (Ω·m·T$^{-1}$) | Carrier Density, n$_e$ (m$^{-3}$) | Mobility, μ$_e$ (cm$^2$·V$^{-1}$·s$^{-1}$) |
|---|---|---|---|
| 50 | $-2.16 \times 10^{-8}$ | $2.89 \times 10^{26}$ | 36.92 |
| 100 | $-1.57 \times 10^{-8}$ | $3.98 \times 10^{26}$ | 23.61 |
| 200 | $-7.06 \times 10^{-9}$ | $8.85 \times 10^{26}$ | 8.54 |
| 250 | $-1.21 \times 10^{-9}$ | $5.16 \times 10^{27}$ | 1.31 |

**Figure captions:**

**Fig. 1**. Schematic of the heat-treatment program followed to synthesize $SnBi_2Te_4$ single crystal, with a photograph of the product as an inset.

**Fig. 2**. (a) XRD pattern acquired from a mechanically cleaved crystal flake of the synthesized $SnBi_2Te_4$ single crystal; (b) Rietveld-refined PXRD pattern of the synthesized $SnBi_2Te_4$ single crystal, in which the inset shows the unit cell constructed using VESTA software.

**Fig. 3**. (a) FESEM image of the synthesized $SnBi_2Te_4$ single crystal; (b) EDS pattern of the synthesized $SnBi_2Te_4$ showing its elemental composition; EDS mappings of (c) Bi, (d) Sn, and (e) Te.

**Fig. 4**. Deconvoluted Raman spectrum of the synthesized $SnBi_2Te_4$ at room temperature, with assignments of its Raman modes shown as an inset.

**Fig. 5**. XPS peaks of the synthesized $SnBi_2Te_4$ single crystal: (a) in the Te 3d region, (b) in the Sn 3d region, and (c) in the Bi 4f region.

**Fig. 6**. (a) Fitted resistivity vs. temperature plot for $SnBi_2Te_4$ in the temperature range 250 K–2 K; (b) $\rho_{xy}$ vs. H plots at 50 K, 100 K, 200 K, and 250 K under applied magnetic fields in the range $\pm 5$ T; (c) variations of carrier density (black line) and mobility (red line) with respect to temperature.

**Fig. 7**. (a) MR% vs. H plot of the synthesized $SnBi_2Te_4$ single crystal at 2 K under applied magnetic fields in the range $\pm 12$ T, with the power-law-fitted high-field MR% data shown as an inset; (b) MR% vs. H plots of the synthesized $SnBi_2Te_4$ single crystal at 2 K, 5 K, 10 K, 50 K, 100 K, and 200 K under applied magnetic fields in the range $\pm 12$ T; (c) HLN-fitted low-field ($\pm 0.5$ T) MC data of the synthesized $SnBi_2Te_4$ single crystal.

**Fig. 8**. (a) The k-space path in the first Brillouin zone for the calculation of bulk electronic band structure; (b) bulk electronic band structure without SOC, including the dispersion correction DFT-D3; (c) bulk electronic band structure without SOC, excluding DFT-D3; (d) bulk electronic band structure with SOC, excluding DFT-D3; calculated PDOS (e) without and (f) with SOC.

**Fig. 9**. Evolutions of the Wannier charge center in the z-plane; the Z2 value is 0 in the $k_z=0$ plane, 1 in the $k_z=0.5$ plane.

**Fig. 10**. (a) Chosen k-path for surface states calculation in reciprocal space; (b) the corresponding plane containing the k-path used for surface states spectrum calculations; (c) surface states spectrum showing a Dirac cone at the $\Gamma$ point near energy 0.1 eV.

**References**

1. M. Z. Hasan and C. L. Kane, Rev. Mod. Phys. **82**, 3045 (2010).
2. X.-L. Qi and S.-C. Zhang, Rev. Mod. Phys. **83**, 1057 (2011).
3. S. Rachel, Rep. Prog. Phys. **81**, 116501 (2018).
4. Z. Ren, A. A. Taskin, S. Sasaki, K. Segawa, and Y. Ando, Phys. Rev. B **82**, 241306(R) (2010).
5. R.J. Cava, H. Ji, M.K. Fuccillo, Q.D. Gibson, Y.S. Hor, J. Mater. Chem. C **1**, 3176 (2013).
6. S. Cai, J. Guo, V. A. Sidorov, Y. Zhou, H. Wang, G. Lin, X. Li, Y. Li, K. Yang, A. Li, Q. Wu, J. Hu, S. K. Kushwaha, R. J. Cava, and L. Sun, npj Quant. Mater. **3**, 62 (2018).
7. K. Kuroda, H. Miyahara, M. Ye, S. V. Eremeev, Yu. M. Koroteev, E. E. Krasovskii, E. V. Chulkov, S. Hiramoto, C. Moriyoshi, Y. Kuroiwa, K. Miyamoto, T. Okuda, M. Arita, K. Shimada, H. Namatame, M. Taniguchi, Y. Ueda, and A. Kimura, Phys. Rev. Lett. **108**, 206803 (2012).
8. T. Menshchikova, S. V. Eremeev, E. V. Chulkov, App. Surf. Sci. **267**, 1 (2013).
9. M. G. Vergniory, T. V. Menshchikova, I. V. Silkin, Yu. M. Koroteev, S. V. Eremeev, and E. V. Chulkov, Phys. Rev. B **92**, 045134 (2015).
10. Poonam Rani, Ankush Saxena, Rabia Sultana, Vipin Nagpal, S. S. Islam, S. Patnaik, and V. P. S. Awana, J. Supercond. Nov. Magn. **32**, 3705–3709 (2019).
11. J. Cui, M. Shi, H. Wang, F. Yu, T. Wu, X. Luo, J. Ying, and X. Chen, Phys. Rev. B **99**, 155125 (2019).
12. H. Li, S. Liu, C. Liu, J. Zhang, Y. Xu, R. Yu, Y. Wu, Y. Zhang, and S. Fan, Phys. Chem. Chem. Phys. **22**, 556 (2020).
13. Ankush Saxena, Poonam Rani, Vipin Nagpal, S. Patnaik, I. Felner, and V. P. S. Awana, J. Supercond. Nov. Magn. **33**, 2251 (2020).
14. Y.-C. Zou, Z.-G. Chen, E. Zhang, F. Kong, Y. Lu, L. Wang, J. Drennan, Z. Wang, F. Xiu, K. Cho, and J. Zou, Nano Res. **11**, 696 (2018).
15. Priyanath Mal, Bipul Das, G. Bera, G. R. Turpu, C. V. Tomy, and Pradip Das, J. Mater. Sci.: Mater. Electron. **33**, 1 (2022).
16. Ankush Saxena, M. M. Sharma, Prince Sharma, Yogesh Kumar, Poonam Rani, M. Singh, S. Patnaik, V. P. S. Awana, J. Alloys Compds. **895**, 162553 (2022).
17. A. Marcinkova, J. K. Wang, C. Slavonic, A. H. Nevidomskyy, K. F. Kelly, Y. Filinchuk, and E. Morosan, Phys. Rev. B **88**, 165128 (2013).
18. M. Nurmamat, K. Okamoto, S. Zhu, T. V. Menshchikova, I. P. Rusinov, V. O. Korostelev, K. Miyamoto, T. Okuda, T. Miyashita, X. Wang, Y. Ishida, K. Sumida, E. F. Schwier, M. Ye, Z. S. Aliev, M. B. Babanly, I. R. Amiraslanov, E. V. Chulkov, K. A. Kokh, O. E. Tereshchenko, K. Shimada, S. Shin, and A. Kimura, ACS Nano **14**(**7**), 9059 (2020).
19. Yunlong Li, Chaozhi Huang, Guohua Wang, Jiayuan Hu, Shaofeng Duan, Chenhang Xu, Qi Lu, Qiang Jing, Wentao Zhang, and Dong Qian, Chinese Phys. B **30**, 127901 (2021).
20. W. Afzal, Z. Yue, Z. Li, M. Fuhrer, X. Wang, J. Phys. Chem. Solids **161**, 110489 (2022).
21. W. Zhao, X. Wang, Adv. Phys. X, **7(1)**, 2064230 (2022).
22. H.-T. He, G. Wang, T. Zhang, I.-K. Sou, G. K. L. Wong, J.-N. Wang, H.-Z. Lu, S.-Q. Shen, and F.-C. Zhang, Phys. Rev. Lett. **106**, 166805 (2011).
23. Deepak Sharma, M.M. Sharma, R.S. Meena, V.P.S. Awana, Physica B: Condens. Matter **600**, 412492 (2021).

24. P. Giannozzi, S. Baroni, N. Bonini, M. Calandra, R. Car, C. Cavazzoni, D. Ceresoli, G. L. Chiarotti, M. Cococcioni, I. Dabo, A. Dal Corso, S. De Gironcoli, S. Fabris, G. Fratesi, R. Gebauer, U. Gerstmann, C. Gougoussis, A. Kokalj, M. Lazzeri, L. Martin-Samos, N. Marzari, F. Mauri, R. Mazzarello, S. Paolini, A. Pasquarello, L. Paulatto, C. Sbraccia, S. Scandolo, G. Sclauzero, A. P. Seitsonen, A. Smogunov, P. Umari, and R. M. Wentzcovitch, J. Phys. Condens. Matter **21**, 395502 (2009).
25. P. Giannozzi, O. Andreussi, T. Brumme, O. Bunau, M. Buongiorno Nardelli, M. Calandra, R. Car, C. Cavazzoni, D. Ceresoli, M. Cococcioni, N. Colonna, I. Carnimeo, A. Dal Corso, S. De Gironcoli, P. Delugas, R. A. Distasio, A. Ferretti, A. Floris, G. Fratesi, G. Fugallo, R. Gebauer, U. Gerstmann, F. Giustino, T. Gorni, J. Jia, M. Kawamura, H. Y. Ko, A. Kokalj, E. Küçükbenli, M. Lazzeri, M. Marsili, N. Marzari, F. Mauri, N. L. Nguyen, H. V. Nguyen, A. Otero-De-La-Roza, L. Paulatto, S. Poncé, D. Rocca, R. Sabatini, B. Santra, M. Schlipf, A. P. Seitsonen, A. Smogunov, I. Timrov, T. Thonhauser, P. Umari, N. Vast, X. Wu, and S. Baroni, J. Phys. Condens. Matter **29**, 465901 (2017).
26. S. Grimme, J. Antony, S. Ehrlich, and H. Krieg, J. Chem. Phys. **132**(15), 154104 (2010).
27. A. A. Mostofi, J. R. Yates, G. Pizzi, Y. S. Lee, I. Souza, D. Vanderbilt, N. Marzari, Comput. Phys. Commun. **185**, 2309 (2014).
28. M. P. L. Sancho, J. M. L. Sancho, and J. Rubio, J. Phys. F: Metal Phys. **14**, 1205 (1984).
29. M. P. L. Sancho, J. M. L. Sancho, and J. Rubio, J. Phys. F: Metal Phys. **15**, 851 (1985).
30. F. J. Manjón, S. Gallego-Parra, P. Rodrıguez-Hernandez, A. Muñoz, C. Drasar, V. Muñoz-Sanjose, and O. Oeckler, J. Mater. Chem. C **9**, 6277 (2021).
31. K. M. F. Shahil, M. Z. Hossain, V. Goyal, and A. A. Balandin, J. Appl. Phys. **111**, 054305 (2012).
32. J. F. Moulder, W. F. Stickle, P. E. Sobol, K. D. Bomben, Handbook of X-ray Photoelectron Spectroscopy, Perkin-Elmer Corporation, Physical Electronics Division, 1992.
33. K. Shrestha, M. Chou, D. Graf, H. D. Yang, B. Lorenz, and C. W. Chu, Phys. Rev. B **95**, 195113 (2017).
34. W. Wang, W. Q. Zou, L. He, J. Peng, R. Zhang, X. S. Wu, and F. M. Zhang, J. Phys. D: Appl. Phys. **48**, 205305 (2015).
35. S. Hikami, A. I. Larkin, and Y. Nagaoka, Prog. Theor. Phys. **63**, 707 (1980).
36. J. Chen, H. J. Qin, F. Yang, J. Liu, T. Guan, F. M. Qu, G. H. Zhang, J. R. Shi, X. C. Xie, C. L. Yang, K. H. Wu, Y. Q. Li, and L. Lu, Phys. Rev. Lett. **105**, 176602 (2010).
37. H. Li, H.-W. Wang, Y. Li, H. Zhang, S. Zhang, X.-C. Pan, B. Jia, F. Song, and J. Wang, Nano Lett. **19**, 2450 (2019).
38. Shama, R. K. Gopal, G. Sheet, and Y. Singh, Sci. Rep. **11**, 12618 (2021).
39. V. K. Maurya, M. M. Patidar, A. Dhaka, R. Rawat, V. Ganesan, and R. S. Dhaka, Phys. Rev. B **102**, 144412 (2020).
40. Z. Hou, Y. Wang, G. Xu, X. Zhang, E. Liu, W. Wang, Z. Liu, X. Xi, W. Wang, and G. Wu, Appl. Phys. Lett. **106**, 102102 (2015).
41. G. Xu, W. Wang, X. Zhang, Y. Du, E. Liu, S. Wang, G. Wu, Z. Liu, and X. X. Zhang, Sci. Rep. **4**, 5709 (2014).
42. P. Niknam, S. Jamehbozorgi, M. Rezvani, and V. Izadkhah, Physica E: Low-Dimensional Systems and Nanostructures, **135**, 114937 (2022).
43. J. P. Perdew, K. Burke, and M. Ernzerhof, Phys. Rev. Lett. **77**, 3865 (1996).

44. Y. Hinuma, G. Pizzi, Y. Kumagai, F. Oba, and I. Tanaka, Comput. Mater. Sci. **128**, 140 (2017).

45. M. D. Moghaddam, S. Jamehbozorgi, M. Rezvani, V. Izadkhah, and M. T. Moghim, Physica E: Low-dimensional Systems and Nanostructures **138**, 115077 (2022).

46. M. Mostafavi, S. Tanreh, M. Astaraki, B. Farjah, M. Rasoolidanesh, M. Rezvani, and M. Darvish Ganji, Physica B: Condensed Matter, **626**, 413446 (2022).

47. M. T. Moghim, S. Jamehbozorgi, M. Rezvani, M. Ramezani, Spectrochimica Acta Part A: Molecular and Biomolecular Spectroscopy, **280**, 121488 (2022).

48. M. Rezvani, M. Astaraki, A. Rahmanzadeh, and M. Darvish Ganji, Phys. Chem. Chem. Phys., **23**(32), 17440–17452 (2021).

49. M. Sabet, S. Tanreh, A. Khosravi, M. Astaraki, M. Rezvani, M. Darvish Ganji, Diamond and Related Materials **126**, 109142 (2022).

50. A. A. Soluyanov and D. Vanderbilt, Phys. Rev. B **83**, 035108 (2011).
51. H. C. Po, A. Vishwanath, and H. Watanabe, Nat. Commun. **8**, 50 (2017).
52. J. Kruthoff, J. de Boer, J. van Wezel, C.L. Kane, R.J. Slager, Phys. Rev. X **7**, 041069 (2017).
53. R.J. Slager, A. Mesaros, V. Juričić, J. Zaanen, Nature Phys **9**, 98 (2013).

Fig. 1

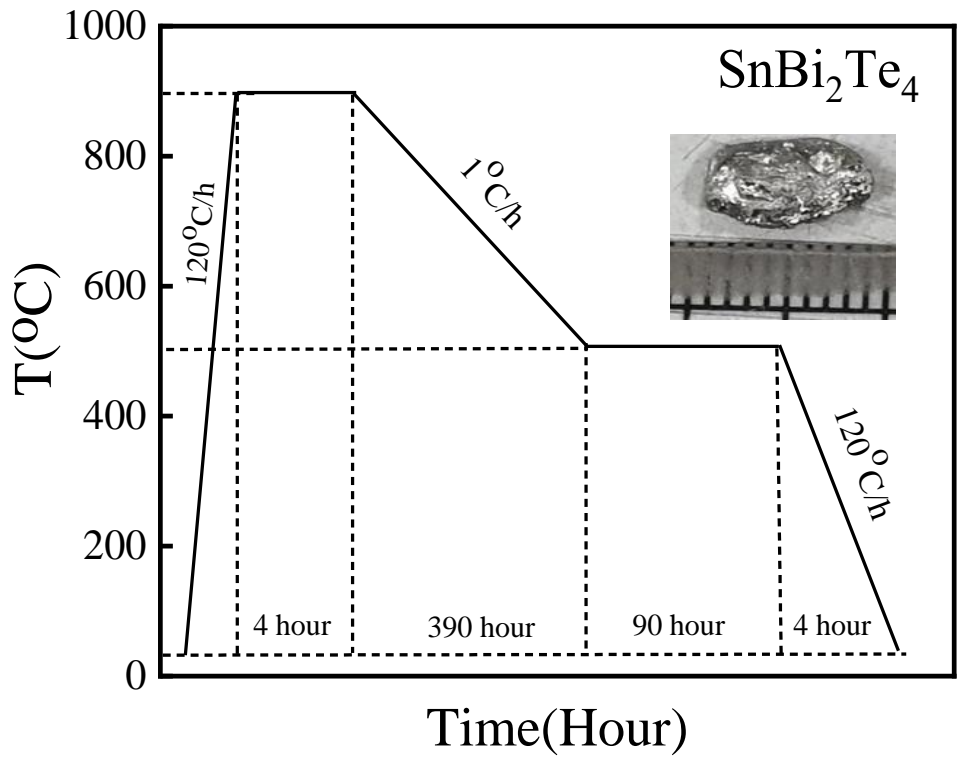

Fig. 2(a)

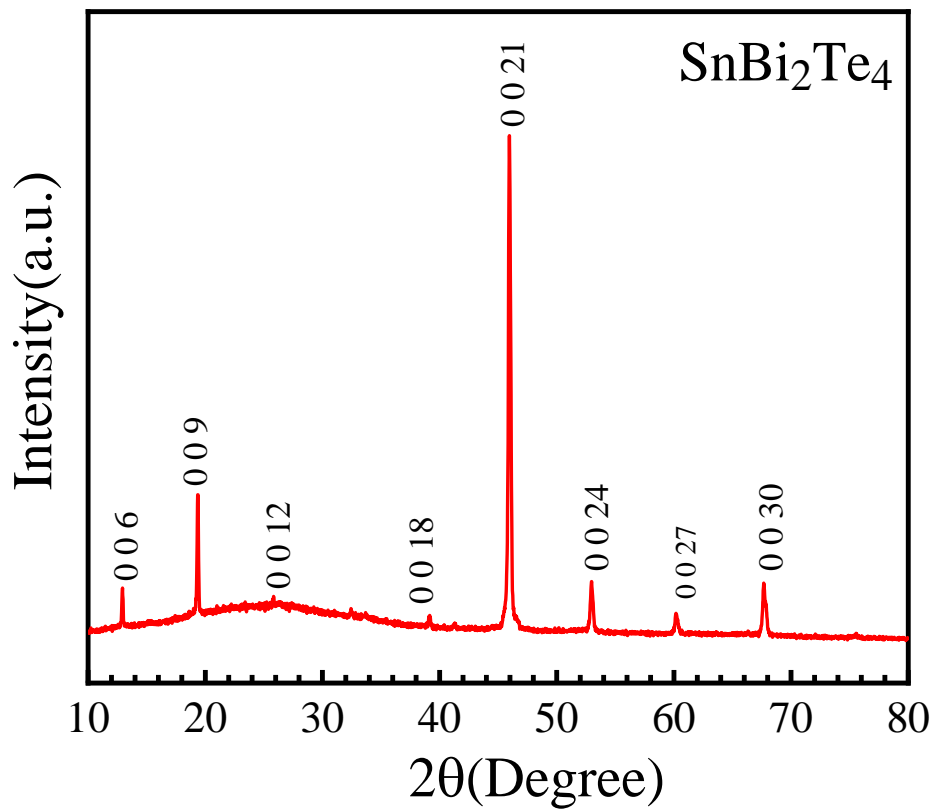

Fig. 2(b)

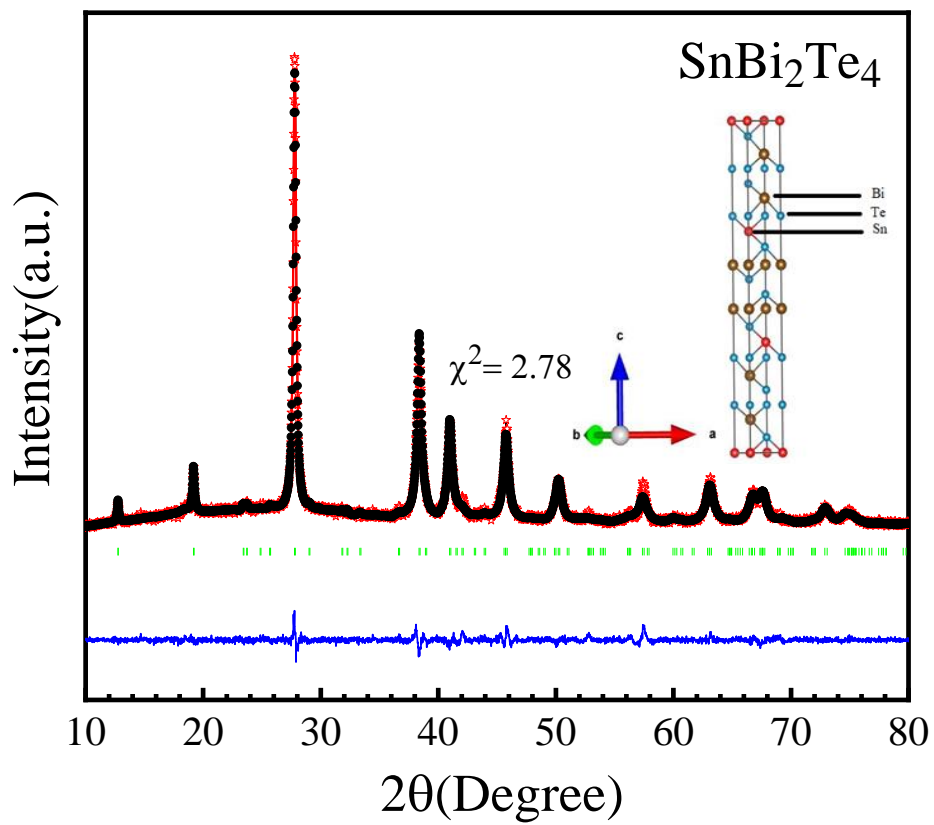

Fig. 3

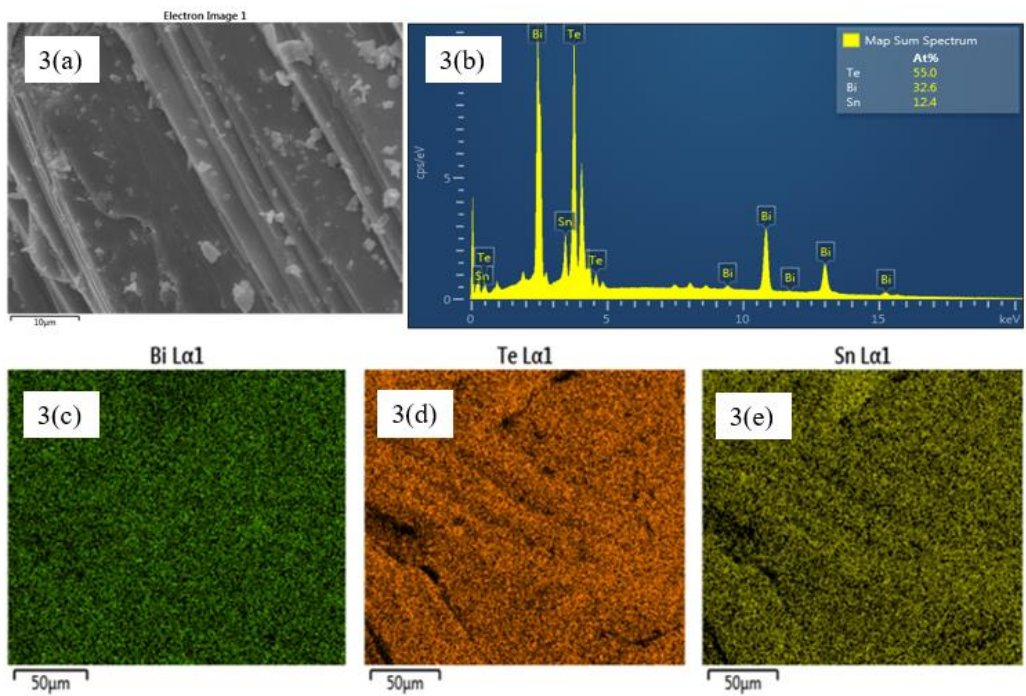

Fig.4

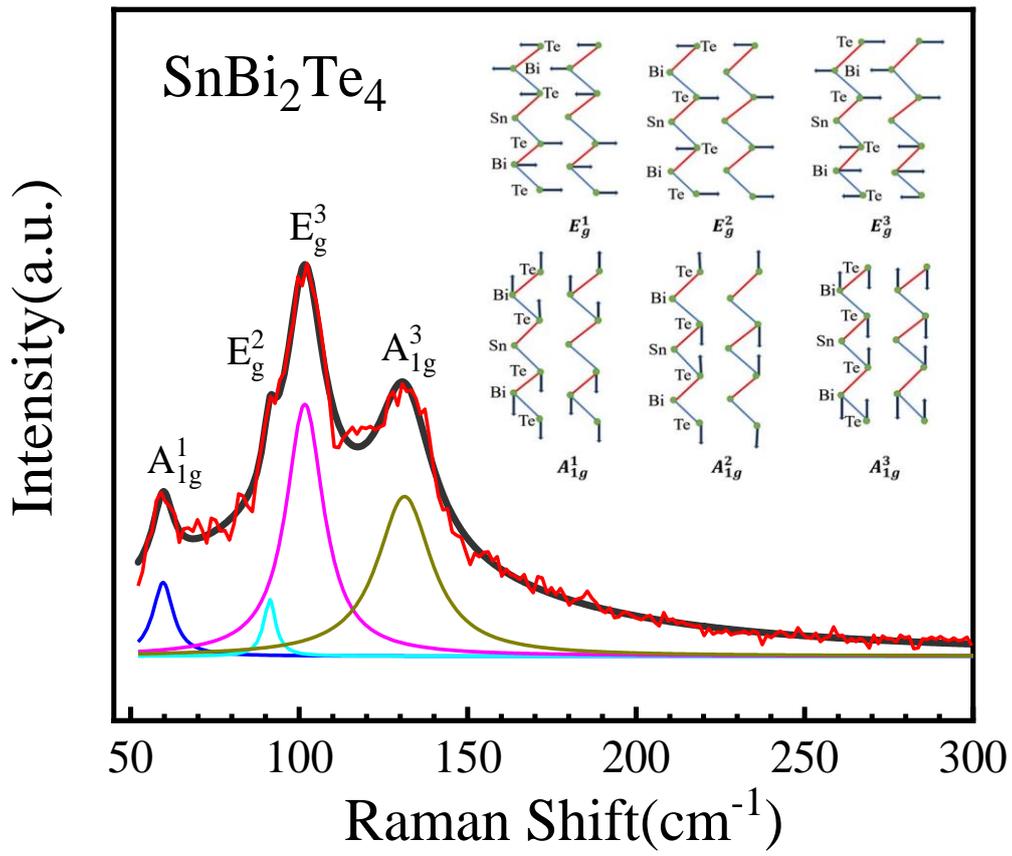

Fig. 5

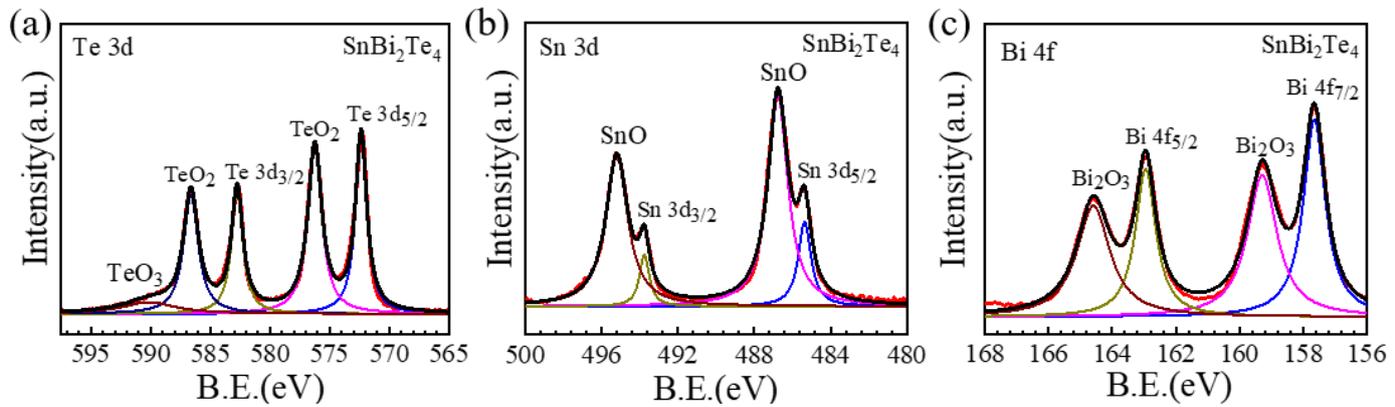

Fig. 6(a)

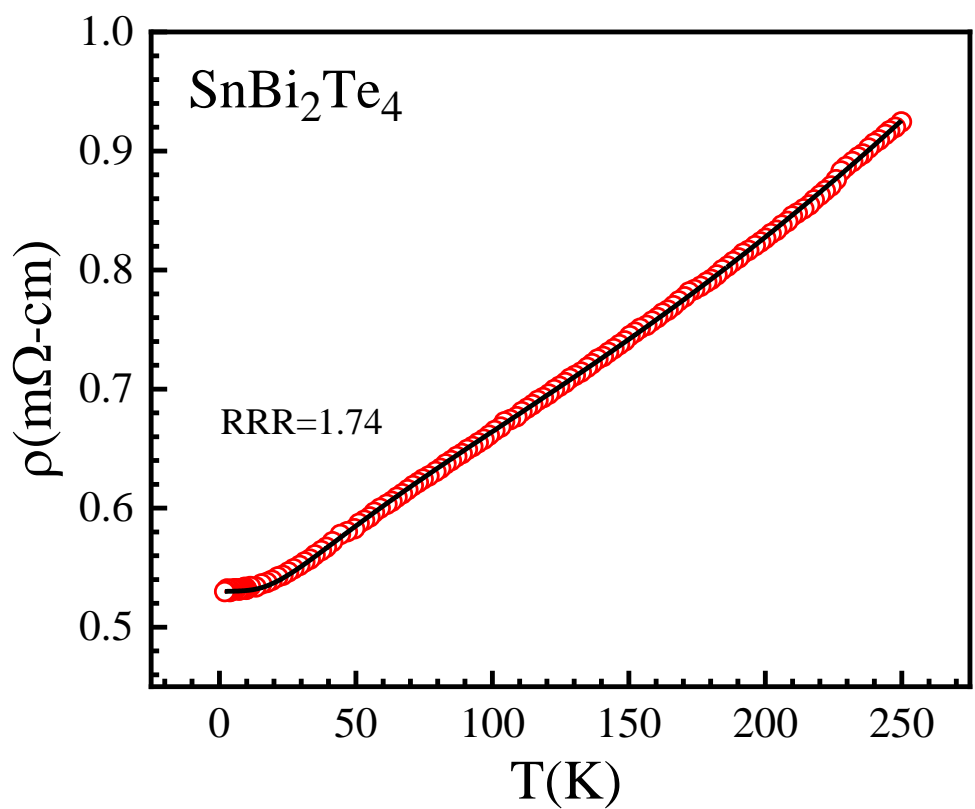

Fig. 6(b)

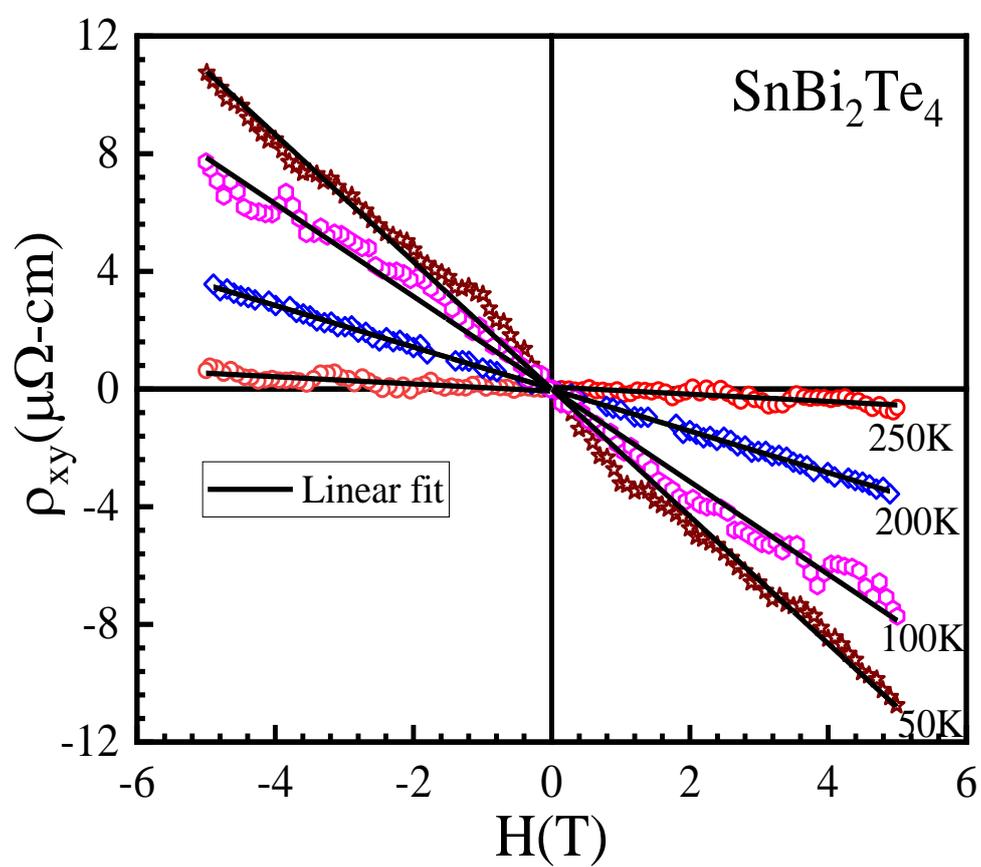

Fig. 6(c)

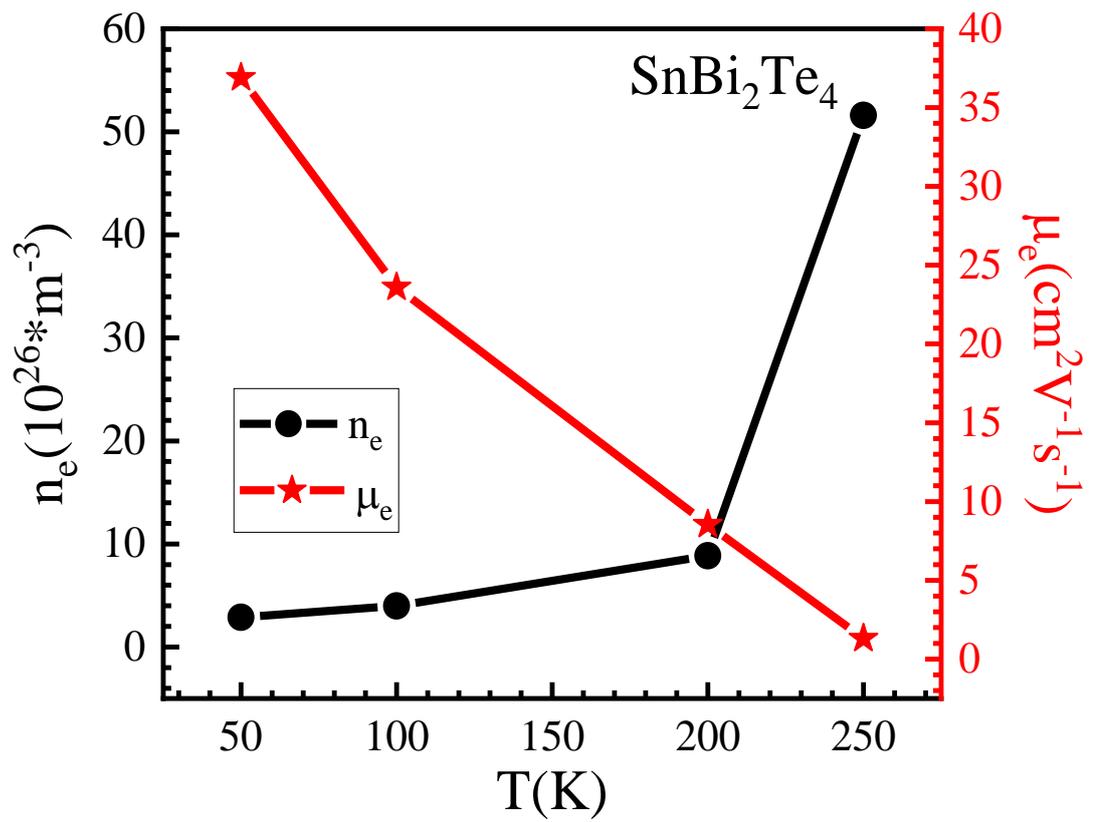

Fig. 7(a)

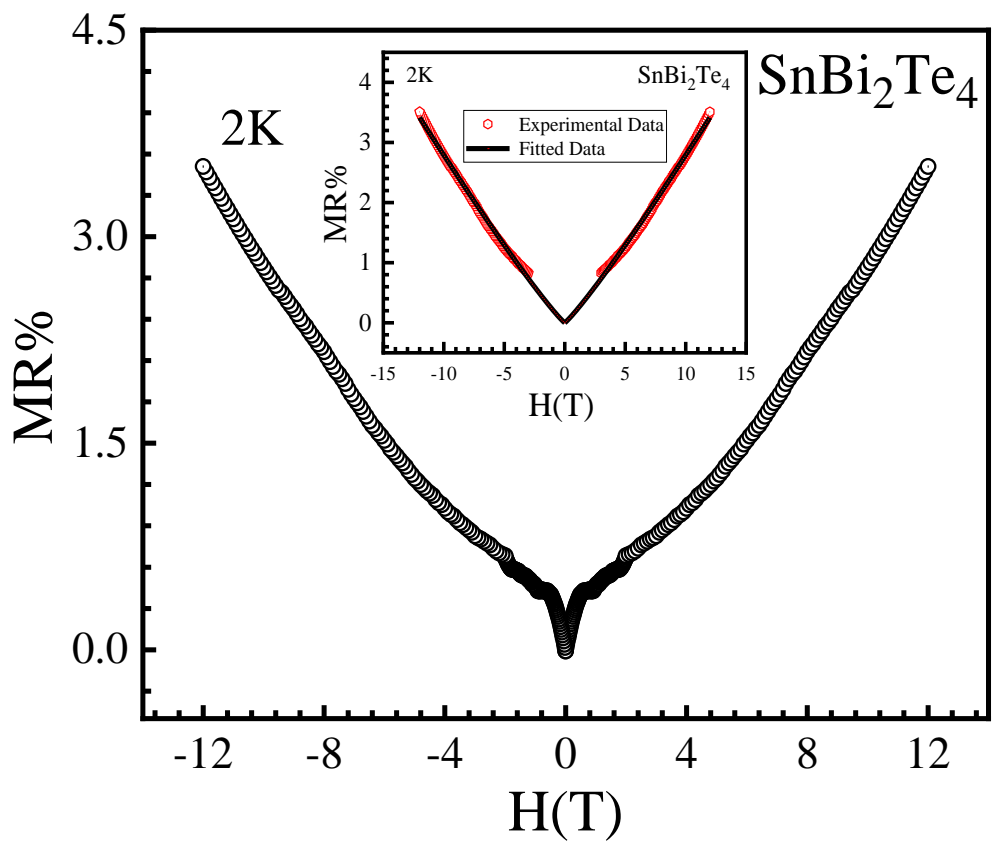

Fig. 7(b).

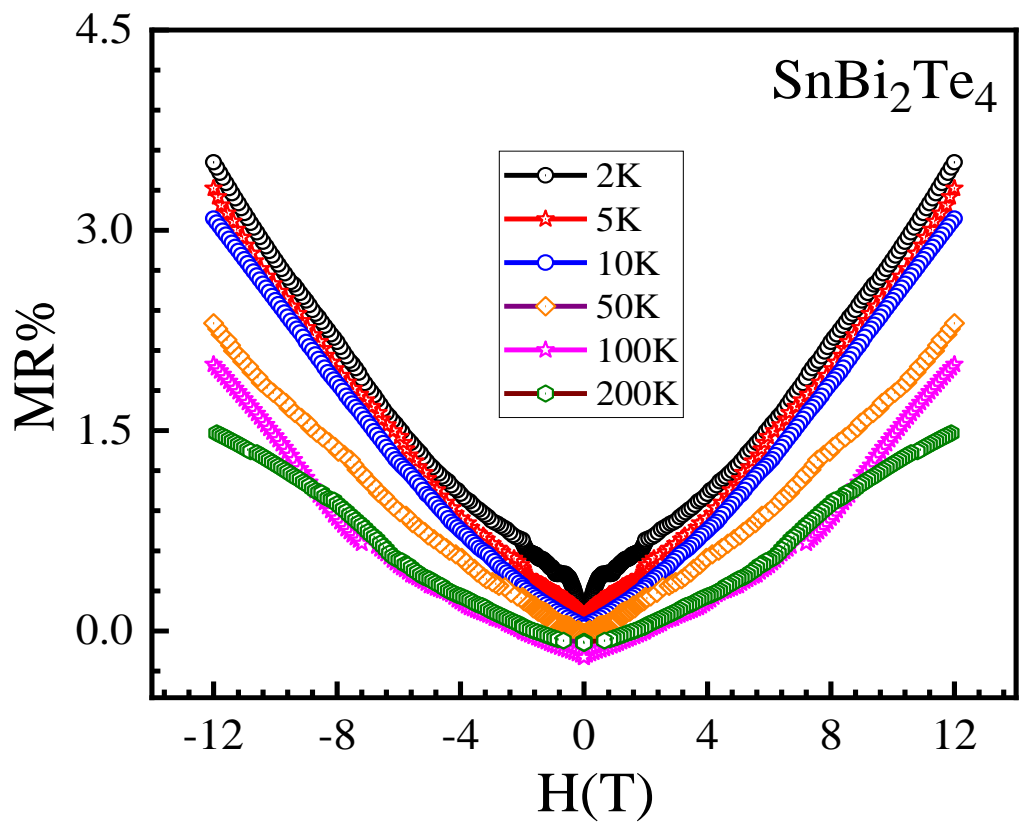

Fig. 7(c)

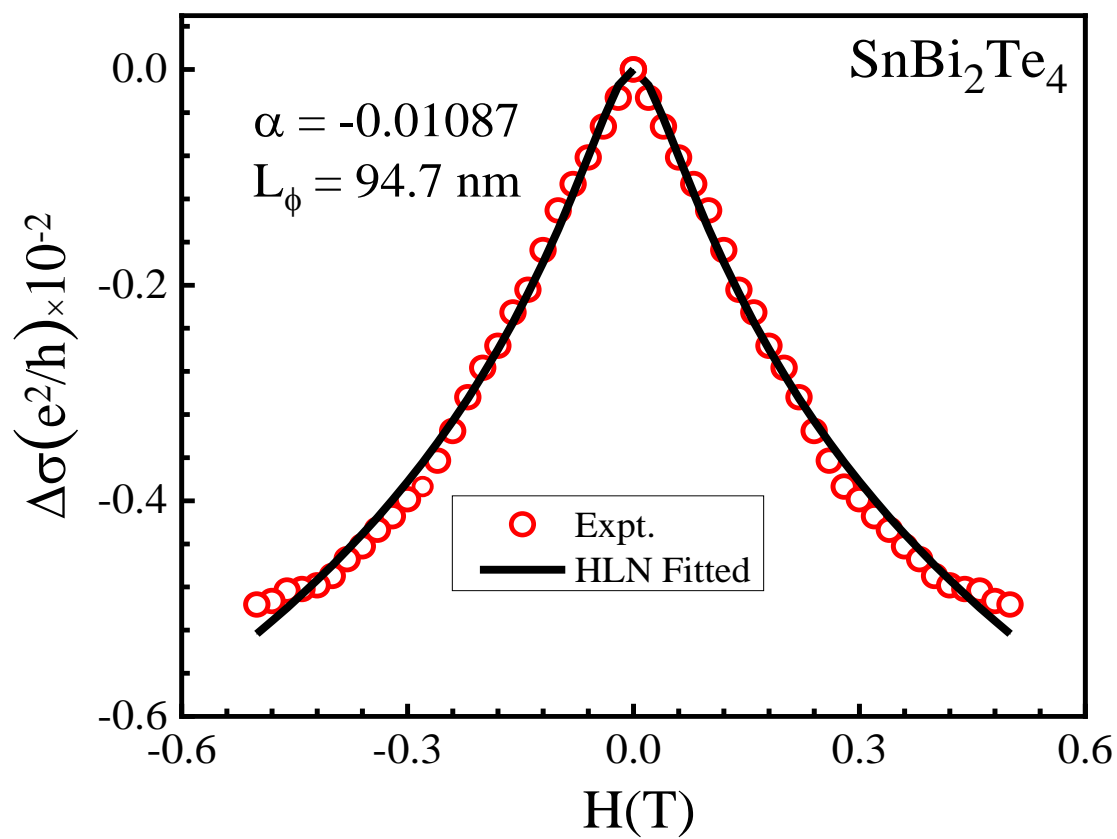

Fig. 8(a)

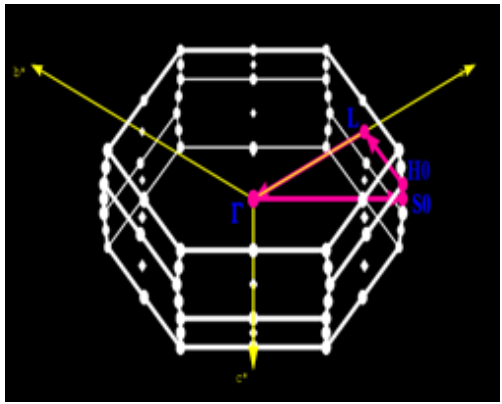

(b)

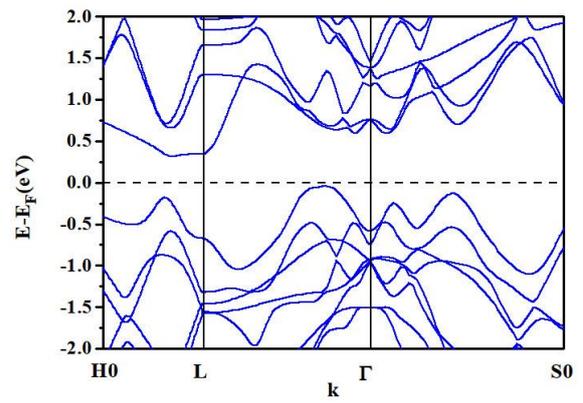

(c)

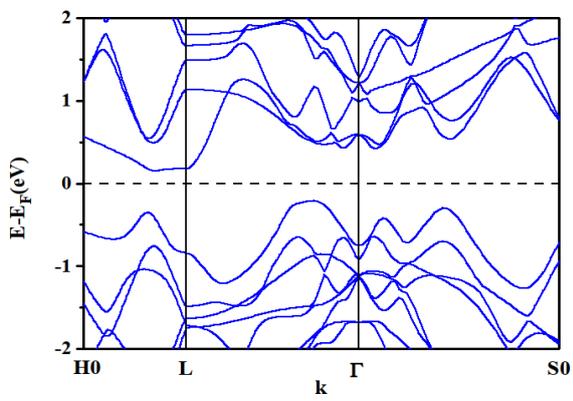

(d)

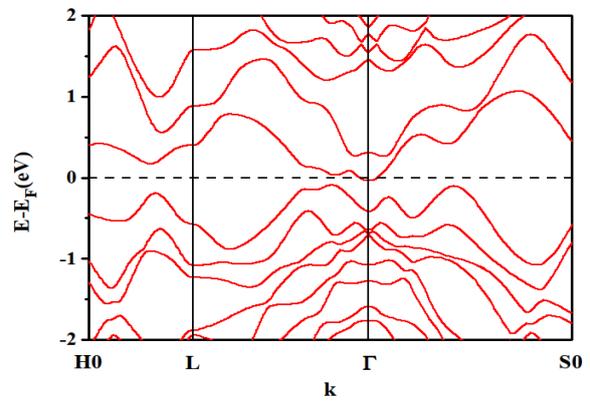

(e)

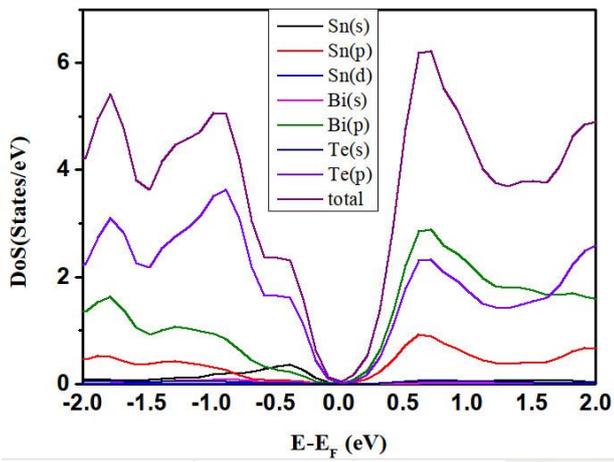

(f)

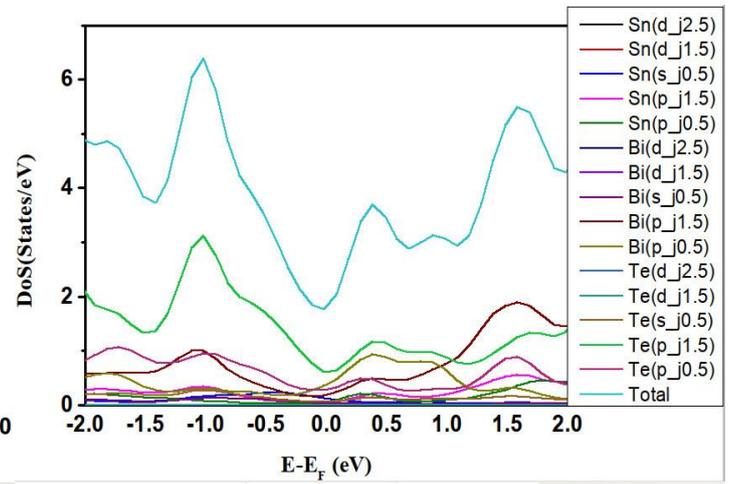

Fig. 9

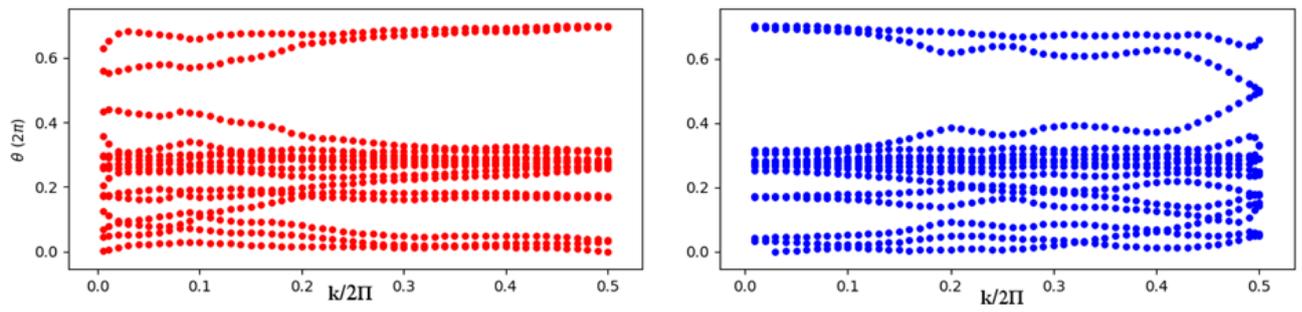

Fig. 10

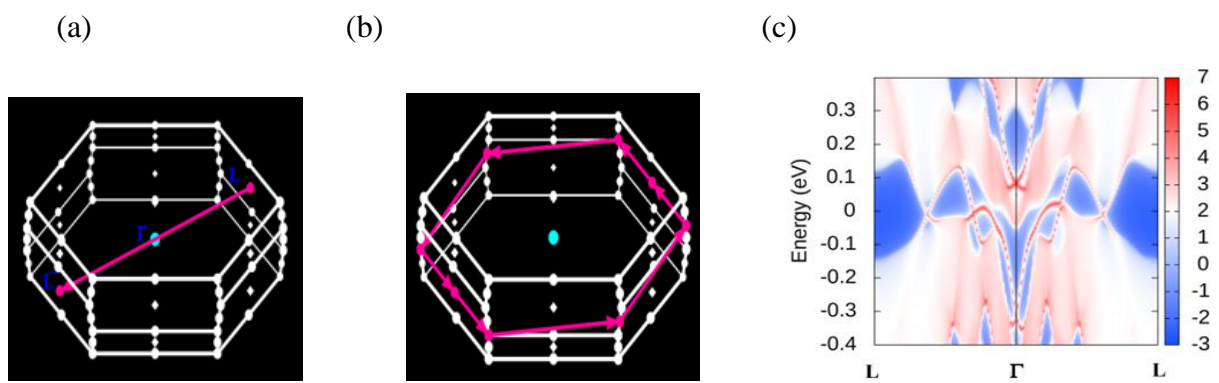